\documentclass[aps,prc,nofootinbib,twocolumn,superscriptaddress,showpacs,
floatfix]{revtex4}
\usepackage{graphicx}% Include figure files
\usepackage{dcolumn}% Align table columns on decimal point
\usepackage{bm}% bold math
\usepackage{amsmath}
\begin{document}

%\graphicspath{{Figure/}}
\title{Hadrons from coalescence plus fragmentation in AA collisions at energies available at the BNL Relativistic Heavy Ion Collider to the CERN Large Hadron Collider }% Force line breaks with \\
\author{Vincenzo Minissale}
 \author{Francesco Scardina}%
 \author{Vincenzo Greco}
%% \email{Second.Author@institution.edu}
 \affiliation{Dipartimento di Fisica e Astronomia, University of Catania, Via S. Sofia 62, 
 I-95123, Catania, Italy.}
 \affiliation{%
 INFN-LNS,Laboratori Nazionali del Sud, Via S. Sofia 64,I-95123, Catania, Italy}


\date{\today}% It is always \today, today,

\begin{abstract}
In a coalescence plus independent fragmentation approach we calculate the $p_T$ spectra of the
main hadrons: $\pi, K, p, \bar p, \Lambda, \phi$ in a wide range of transverse momentum from low $p_T$ up to about 10 GeV.
The approach in its main features was developed several years ago at RHIC energy. Augmenting
the model with the inclusion of some more main resonance decays, we show that the approach
correctly predicts the evolution of the $p_T$ spectra from RHIC to LHC energy and in particular
the baryon-to-meson ratios $p/\pi,\bar p/\pi,\Lambda/K$ that reach a value of the order of unit at $p_T \sim 3\, \rm GeV$.
This is achieved without any change of the coalescence parameters. The more recent availability
of experimental data up to $p_T \sim 10\rm\, GeV$ for $\Lambda$ spectrum as well as for $p/\pi$ 
and $\Lambda/K$ shows some lack of yield in a limited $p_T$ range around 6 GeV.
This indicates that the baryons $p_T$ spectra from AKK fragmentation functions are too 
flat at $p_T\lesssim 8 \,\rm GeV$. We also show that in a coalescence plus fragmentation approach 
one predicts a nearly $p_T$ independent $p/\phi$ ratio up to $p_T \sim 4 \rm\, GeV$ followed by
a significant decrease at higher $p_T$. Such a behavior is driven
by a similar radial flow effect at $p_T< 2 \,\rm GeV$ and the dominance of fragmentation for $\phi$
at larger $p_T$.

\end{abstract}

\maketitle

%\tableofcontents
\section{Introduction}
Heavy-Ion collisions at ultra-relativistic energies generates the creation of a new state of matter
with a temperature $T$ significantly above the cross-over temperature $T_c\simeq 160 \rm MeV$
expected for the transition of hadronic matter to the Quark-Gluon Plasma (QGP) \cite{Aoki:2006we}.
In the last decade a consensus seems to have emerged that quarks and gluons are indeed deconfined 
for a short amount of time in the fireball created, and that this quark gluon plasma 
behaves like a very good liquid with small viscosity over entropy density ratio $\eta/s$ 
\cite{Shuryak:2008eq,Jacak:2012dx,Heinz:2013th}.
One of the most surprising observations at RHIC \cite{Adare:2013esx} have been the strong enhancement of the 
baryon over meson ratio in a wide range of  $3 \rm \, GeV \lesssim p_T \lesssim 6 \,\rm GeV$ and
a significantly larger (about $40-50\%$ ) elliptic flow $v_2$ of baryons with respect to that of mesons \cite{Adams:2004bi,Adams:2005zg,Adler:2003kt},
at variance with expectation from both hydrodynamics and jet-quenching fragmentation.
However soon after these first data it was realized that an hadronization process
through quark coalescence can naturally explain both features of hadron spectra appearing in the
same $p_T$ region \cite{Fries:2003vb,Greco:2003xt,Greco:2003mm,Fries:2003kq,Hwa:2002tu,Hwa:2004ng,Molnar:2003ff,Fries:2008hs}. An early extrapolation at LHC energy for the $p/\pi$ ratio
has been performed already a decade ago \cite{Fries:2003fr}.

The idea of quark coalescence as hadronization from a QGP where the quark (anti-quark) needed 
to hadronize are already present and there is no need to create them from vacuum, was initially
suggested in Refs. \cite{Biro:1994mp,Biro:1998dm} with the approach known as ALCOR  model. 
However, there the main focus was on particle yields and ratios and not the different kinematics
between baryons and mesons that explains their pattern for $p_T$ spectra and elliptic flow.
%\emph{On the other hand ALCOR forcing algebraically unitarity has shown for the first time that if
%coalescence into resonances is taken into account than one can explain satisfactorily
%the yields of all the hadrons observed in relativistic heavy-ion collisions.}

After the first papers applying the coalescence mechanism for hadron production at RHIC 
\cite{Fries:2003vb,Greco:2003xt,Greco:2003mm,Fries:2003kq,Molnar:2003ff}
there have been mainly two further studies.In one case the coalescence approach has been
extended to include finite width that accounts for off-shell effect which allows to
include the constraint of energy conservation \cite{Ravagli:2007xx,Ravagli:2008rt}. This is seen
not to change the general feature of the coalescence mechanism, the baryon/meson enhancement and approximate 
quark number scaling of $v_2$, but shows that at low $p_T$, when energy conservation plays a stronger
role, a scaling with $m_T=(p_T^2+m^2)^{1/2}$ is favored, at least for direct hadron production.
More recently, an approach that from one hand enforce algebraically the unitarity similarly
to ALCOR approach, but for the momentum distribution use essentially the prescription of Refs.
\cite{Greco:2003xt,Greco:2003mm} has been developed in \cite{Shao:2009uk,Wang:2012cw}.
The price paid to have both aspects is the need to add a global constant in the
coalescence probability in such a way that all the quarks recombine and moreover
the quark distribution functions are fitted by mean of six parameters to describe hadron spectra. Nonetheless
the result are similar to \cite{Greco:2003xt,Greco:2003mm} except a better description
of the low $p_T$ region, where however the present coalescence scheme can be anyway questionable
and the success of the description of the spectra down to low $p_T$ may simply come from
the large flexibility of the fitted parton distribution.

In this paper, we employ the coalescence approach developed in 
\cite{Greco:2003mm,Greco:2007nu} that is based on the same basic idea
of phase-space quark coalescence as in \cite{Fries:2003kq} but being solved by a Monte Carlo
approach allows to include a 3D geometry as well as radial flow correlation in the partonic
spectra and the effect of the main resonance decays. The last is mainly discussed in \cite{Fries:2008hs} and allows
to have a reasonable description of the spectra also at low momenta that are usually dominated
by feed-down from resonance decays. However the model does not
implement a unitarity conservation that guarantees hadronization of the full bulk of particles,
also energy conservation and the finite width (off-shell) effect are not included as in 
\cite{Ravagli:2007xx,Ravagli:2008rt} which at low momentum play a role.
Therefore the model has to be considered really applicable only at $p_T \gtrsim 1.5 \rm\, GeV$,
even if the description of the spectra or baryon/meson ratios remains approximately
good also at lower $p_T$, especially for pions that are dominated by decay feed-down
at low $p_T$. In fact the present dynamical formulation of the coalescence used remains fully meaningful when the single process of hadronization is small and hence it is not justified where the bulk of the particles resides. 

We report in this paper the results applying the original model of coalescence plus fragmentation 
in \cite{Greco:2003mm} at both LHC Pb+Pb $\sqrt{s}=2.76\,\rm ATeV$ and RHIC Au+Au $\sqrt{s}=200\,\rm AGeV$ energies. This last case was partially present also in \cite{Greco:2003mm}, but
some more resonance decays have been added, the update Albino-Kniehl-Kramer (AKK) fragmentation
function and also the $\Lambda$ momentum spectra have been evaluated.
Furthermore the availability of experimental data allows for a comparison in a wider $p_T$ range.
We see that the same model is able to correctly predict the spectra of the main hadrons ($\pi,K,p,\bar p,\Lambda$),
and correctly account for the evolution of such spectra from RHIC to LHC energy and in particular the $p_T$ dependence of the $p/\pi,\bar p/\pi,\Lambda/K_s^0$ ratios.

The paper is organized as follows: in Sec. II we describe the general formalism of a covariant coalescence model for mesons and baryons and how it is numerically solved by means of a Monte Carlo method. How mini-jets and the quark-gluon plasma partons are determined is described in Sec. III. Results for the transverse momentum hadronic spectra obtained from the coalescence model are given in Sec. IV for RHIC energy and in Sec. V for LHC energy.
Finally, we conclude in Sec. VI with a summary of present work and an outlook about
future developments and applications of the parton coalescence model.

\section{Coalescence plus Fragmentation model}

The hadronization mechanism is well known to belong to the non-perturbative domain of the QCD
dynamics and certainly a first-principle description
of hadron formation has yet to be obtained. One of the most common approach dealing
with hadronization, routinely used in nuclear and particle physics for inclusive hadron production,
is the independent fragmentation in which a single colored parton $ a $ has to hadronize into the
hadron $h$. For this purpose, fragmentation or parton decay functions $D_a(z) \rightarrow h$ have been defined and give the probability of finding hadron $ h$ in parton $a$ with a momentum fraction $z$,
$0 < z < 1$. The cross section for inclusive hadron production can then be written as
$\sigma_h= \sigma_a \otimes D_{a\rightarrow h}$
which is a convolution of the production cross section $\sigma_a$ for parton $ a$ with the fragmentation function $D_{a \rightarrow h}(z)$.
Fragmentation functions are not reliably calculable from first principles
in QCD. However, they are observables and can be measured experimentally.

Physically, the
fragmentation of a single parton happens through the creation of $\bar q q$ pairs, 
which subsequently arrange into color singlets forming hadrons.
Such a scheme is based on the concept of QCD factorization,
which separates the long- from the short-distance dynamics. 
Notice that in a bulk QGP matter created in heavy-ion collisions the hadronization can occur
on average at a time scale $t_h \sim 4-10 \, \rm fm/c$ (depending on its initial temperature)
this means that the partons do not need a creation of $q \bar q $ at the moment
of hadronization, but they can rescatter and 
generate a thermal medium expanding with a radial collective flow. At $t_h$ 
partons have a certain abundance in phase space such that there is no
need for the creation of additional partons through splitting or string breaking. The most naive
expectation for such a scenario is a simple recombination of the deconfined partons into hadrons.
Indeed, there is experimental evidence that this is the correct description of hadronization,
even long before a thermal occupation of parton phase space is reached \cite{Adamovich:1993kc,Aitala:1996hf,Braaten:2002yt,Rapp:2003wn}.

\subsection{Coalescence}
The approaches developed in \cite{Greco:2003mm,Greco:2003vf,Fries:2003kq,Fries:2003vb}
for the coalescence is based on the Wigner formalism originally developed for nucleons \cite{Dover:1991zn}.  
The spectrum of hadrons formed from the coalescence of quarks can be written as:
\begin{eqnarray}
\label{eq-coal}
\frac{d^{2}N_{H}}{dP_{T}^{2}}&=& g_{H} \int \prod^{n}_{i=1} \frac{d^{3}p_{i}}{(2\pi)^{3}E_{i}} p_{i} \cdot d\sigma_{i}  \; f_{q_i}(x_{i}, p_{i})\nonumber \\ 
&\times& f_{H}(x_{1}..x_{n}, p_{1}..p_{n})\, \delta^{(2)} \left(P_{T}-\sum^{n}_{i=1} p_{T,i} \right)
\end{eqnarray}

where $d\sigma_{i}$ denotes an element of a space-like hypersurface, $g_{H}$ is the statistical factor to form a colorless hadron from quark and antiquark with spin 1/2. $f_{q_i}$ are the quark (anti-quark) distribution in phase space. $f_{H}$ is the Wigner function and describes the spatial and momentum distribution of quarks in a hadron. 
Eq. \ref{eq-coal} for $i=2$ describes meson formation, and for $i=3$ the baryon one.
The above formula can also be used for antibaryons by replacing quark
momentum spectra by the momentum spectra of antiquarks

The statistical factor $g_H$ takes into account the internal quantum numbers in forming
a colorless hadron from spin-1/2 colored quark and antiquark. For
mesons considered here, i.e., $\pi, \rho ,\omega, K$, and $K^*$, the statistical
factors are $g_\pi=g_K=1/36$ and $g_\rho=g_{K^*}=1/12$. 
For baryons and antibaryons considered
in present study, i.e. $p, \Delta, \bar p$ and $\bar\Delta,\Lambda $, the statistical
factors are $g_p=g_{\bar p}=g_\Lambda=1/108$ and $g_\Delta=g_{\bar \Delta}=1/54$. 
The coalescence probability function $f_M(x_1,x_2;p_1,p_2)$  ($f_B(x_1,x_2,x_3;p_1,p_2,p_3)$)depends in principle on the
overlap of the quark and anti-quark distribution functions with the
wave function of the meson (baryon) as well as the interactions of emitted virtual partons, which are needed for balancing the energy and momentum, with the partonic matter. Neglecting the off-shell effects the coalescence probability
function is then simply the covariant hadron Wigner distribution function.

In the Greco-Ko-Levai (GKL) approach in Ref.\cite{Greco:2003mm} for a meson the Wigner function is a sphere of radius $\Delta_{x}$ in coordinate space and $\Delta_{p}$ in momentum space, these two parameters are related by the uncertainty principle $\Delta_{x}\Delta_{p}=1$,  
\begin{eqnarray} 
f_{M}(x_{1}, x_{2}; p_{1}, p_{2}) &=& \frac{9\,\pi}{2}\,
%{2(\Delta_{x}\Delta_{p})^{3}}
\; \Theta\left(\Delta_{x}^{2}-x_{r}^{2}\right) \nonumber \\
&\times & \Theta \left(\Delta_{p}^{2} - p_{r}^{2}+\Delta m_{12}^{2}\right) \nonumber
\end{eqnarray}
where we have defined the quadri-vectors for the relative coordinates
$x_{r1}=x_{1} - x_{2}$, $p_{r1}=p_1-p_2$ and the scalar $\Delta m_{12}=m_1-m_2$.
For a baryon we have a similar Wigner function expressed in term of appropriate relative coordinates:
\begin{eqnarray}
f_B&=&
%(x_1,x_2,x_3;p_1,p_2,p_3)=
\frac{81\,\pi^2}{4} %\nonumber\\
\Theta\left(\Delta_x^2-\frac{1}{2} x_{r1}^2\right)
\times\Theta\left(\Delta_p^2-\frac{1}{2} p_{r1}^2\right) \nonumber\\
&&\Theta\left(\Delta_x^2-x_{r2}^2\right)
\times\Theta\left(\Delta_p^2-p_{r2}^2-\Delta m_{123}^2\right),
\end{eqnarray} 
and we have defined $x_{r2}=(x_1+x_2-2x_3)/\sqrt{6}$ and $p_{r2}=(p_1+p_2-2p_3)/\sqrt{6}$
and $\Delta m_{123}= m_1+m_2-2m_3$. We note that
the Wigner function for the meson has only one parameter $\Delta_p=\Delta_x^{-1}$ that we choose to have a mean square radius of mesons of about 0.8 fm which corresponds to a $\Delta_{p}^{meson}=0.19 \, \rm GeV$.
For baryons we have also one parameter, even if one could consider two different $\Delta_p's$ for the two relative coordinates, for $qqq$ baryons we have a $\Delta_{p}^{proton}=0.33 \,\rm GeV$ and for $qqs$ baryons $\Delta_{p}^{\Lambda}=0.38 \,\rm GeV$ which corresponds to a slighter smaller radius
for $\Lambda$ as one can expect
from the wave function of an harmonic oscillator that scales with the inverse square root of the reduced
mass of the system. However we have checked that small variation of $\Delta_p$ does not strongly affect the slope of the spectra coming from coalescence.

A specific feature of the GKL approach has been the inclusion of resonance decays to the pion, proton, 
$K$ spectra. This was a way to include a minimal effect that certainly is present in the final spectra
observed. In this paper we have also included some more resonance with respect to the original work 
\cite{Greco:2003mm} this however does not change significantly the results for the $p_T$ spectra
especially at $p_T\gtrsim 1.5 \rm \,GeV$. This can be envisaged because the resonance decay products mainly feed-down the low
$p_T$ region. The residual effect at intermediate $p_T$ can be reabsorbed by a rescaling of the
Wigner function hadron parameter $\Delta_p$ by about a $10-15\%$.
For the resonances the coalescence probability is augmented with a suppression factor
that takes into account for the Boltzmann probability to populate an excited state of energy 
$\Delta \, E$ at a temperature T, namely accordingly to the statistical model factor
$exp(-\Delta E/T)$, with $\Delta \,E=E_{H^*}-E_H$ and $E_{H^\star}=(p_T+m_H^\star)^{1/2}$ and
$m_H^\star$ the mass of the resonance;
of course also the degeneracy factors that come from the different values of isospin and total angular momentum are taken in account.

\subsection{Fragmentation} 

In Ref.\cite{Fries:2003kq,Greco:2003mm,Fries:2008hs} it has been clarified that at increasing $p_T$ the probability to coalescence decreases and eventually the standard independent fragmentation takes over.
It is therefore necessary to include also the contribution from the mini-jet fragmentation.
This is done by employing the same parton distribution function
that at high $p_T>p_0\sim 3\,\rm GeV$ are those that can be calculated in next-to-leading order (NLO) in a pQCD scheme. However in AA collisions one must include also the modification due to
the jet quenching mechanism \cite{Gyulassy:2003mc,Wang:1998bha}. The hadron momentum spectra from the mini-jet parton spectra is given by:
\begin{equation}
\frac{dN_{had}}{d^{2}p_T\,dy}=\sum_{jet}\int dz \frac{dN_{jet}}{d^{2}p_T\, dy} \frac{D_{had/jet}(z,Q^{2})}{z^{2}} 
\end{equation}

where $z=p_{had}/p_{jet}$ is the fraction of mini-jet momentum carried by the hadron and $Q^2=(p_{had}/2z)^2$ is the momentum scale for the fragmentation process. For $D_{had/jet}(z,Q^{2})$ we employ
the AKK fragmentation function \cite{Albino:2005me,Albino:2008fy}. 
The fragmentation is applied for partons at $p_T> 3\, \rm GeV$.
It is also worth to mention that the inclusion of both coalescence and fragmentation
does not lead to a double counting of hadron formed because a hadron at $p_T^h$
from coalescence comes from partons at $p_T \simeq p_T^h/n$ with $n=2,3$, while
from fragmentation the parton creating it comes
from a $p_T \simeq 1.5-2\,p_T^h$. Therefore, for example considering the fragmentation for a
hadron at $p_T^h\simeq \, 3 \,\rm GeV$ the quark that creates it has on average a
$p_T\approx 5\,\rm GeV$ but the probability of coalescence for such a parton
is very small and it would produce
hadrons in a very different region of $p_T^h \gtrsim 10-15\,\rm GeV$ where
the coalescence is certainly negligible.

As in \cite{Greco:2003mm} the multi-dimensional integrals in the coalescence formula
as well as the fragmentation  are evaluated by the 
Monte-Carlo method via test particles. Specifically, we introduce a large 
number of test partons with uniform momentum distribution in phase-space. To take into 
account the large difference between numbers of thermal and mini-jet 
partons, a test parton with momentum ${\bf p}_{\rm T}$ is given a 
probability that is proportional to the parton momentum distribution, 
e.g., ${dN_q}/{d^2{\bf p}_{\rm T}}$, with the proportional 
constant determined by requiring that the sum of all parton probabilities 
is equal to the parton number. With test partons, the coalescence formulas, 
for mesons and baryons can be re-written as
\begin{eqnarray}
\frac{dN_M}{d^2{\bf p}_{\rm T}}&=&g_M
\sum_{i,j}P_q(i)P_{\bar q}(j)\delta^{(2)}({\bf p}_{\rm T}-{\bf p}_{i{\rm T}}
-{\bf p}_{j{\rm T}})\nonumber\\
&\times&f_M(x_i,x_j;p_i,p_j).
\end{eqnarray}
and
\begin{eqnarray}
\frac{dN_B}{d^2{\bf p}_{\rm T}}&=&g_B
\sum_{i\ne j\ne k}P_q(i)P_q(j)P_q(k)\nonumber\\
&\times&\delta^{(2)}({\bf p}_{\rm T}-{\bf p}_{i{\rm T}}-{\bf p}_{j{\rm T}}-
{\bf p}_{k{\rm T}})\nonumber\\
&\times&f_B(x_i,x_j,x_k;p_i,p_j,p_k).
\end{eqnarray}
In the above, $P_q(i)$ and $P_{\bar q}(j)$ are probabilities carried by 
$i$th test quark and $j$th test antiquark. 

The Monte-Carlo method allows us to treat the coalescence 
of low momentum partons on the same footing as that of high momentum ones.

\section{Fireball and parton distributions}
The coalescence approach as developed till now to describe hadron production
in uRHIC's is based on a fireball where the bulk of particles is a thermalized system
of gluons and $u,d,s$ quarks and anti-quarks
at the temperature $T_c=165 \rm \, MeV$ \cite{Aoki:2006we} which is about the temperature for the cross-over transition in realistic lattice QCD calculation \cite{Borsanyi:2010cj}. At high momenta, $p_T> 2.5 \rm \, GeV$,
the distribution is taken to be the partonic spectra that undergone the in-medium
jet quenching \cite{Scardina:2010zz}.
The longitudinal momentum distribution is assumed to be boost-invariant, i.e., a uniform rapidity distribution in the range $y\in(-0.5,+0.5)$.
While these features are common to both RHIC and LHC, with the beam energy there is a change
in the radial flow and self-consistently in the volume of the bulk of the QGP, as well as
at high $p_T$ there is a different initial distribution and jet quenching.

%\subsection{Bulk QGP}
In this paper, we consider 
Au+Au at $\sqrt{s_{NN}}=200$ GeV at RHIC and Pb+Pb collision at $\sqrt{s_{NN}}=2.76$ TeV at the Large Hadron Collider (LHC) for the central collisions ($0-10\%$).
To take into account for the quark-gluon plasma collective flow, we assume for the partons a  velocity linear radial profile as $\beta_T=\beta_{max}\frac{r}{R}$, where $R$ is the transverse radius of the fireball. 
The quark and antiquark distribution up to 2 GeV are hence given by a thermal distribution with flow:
\begin{equation}
\label{quark-distr}
\frac{dNq,\bar{q}}{d^{2}r_{T}\:d^{2}p_{T}} = \frac{g_{q,\bar{q}} \tau m_{T}}{(2\pi)^{3}} \exp \left(-\frac{\gamma_{T}(m_{T}-p_{T}\cdot \beta_{T} \mp \mu_{q})}{T} \right) 
\end{equation}
here $g_{q}=g_{\bar{q}}=6$ indicates the spin-color degeneracy of light
quarks and antiquarks, and the minus and plus signs are
for quarks and antiquarks, respectively. The temperature is set to  $T=165\text{ MeV}$. 
Eq. (\ref{quark-distr}) also applies to gluons 
after replacing $g_q$ by the gluon spin-color degeneracy $g_g=16$ and 
dropping the chemical potential. For the gluon mass, we take it
to be similar to that of light quarks in order to take into account
non-perturbative effects in the quark-gluon plasma. Because $m_g<2\,m_q$,
we convert $gg\rightarrow q\bar q$. 
The longitudinal 
positions of partons are then determined by $z=\tau\sinh y$, as we have assumed Bjorken correlation
$\eta=y$ at particle level which implies neglecting the longitudinal thermal motion. Such an
approximation should not affect the slope of the $p_T$ spectra.

The radial flow $\beta_{max}$ and the volume $V=\pi R^2_\perp \tau$ (in one unit of rapidity) could be in general considered as parameters evaluated accordingly to the typical value of lifetime of the QGP and, assuming a constant acceleration, we connect the radial expansion with the radial flow $R_\perp=R_0+0.5 \beta_{max} \tau$.
So the radial flow and the volume are constrained imposing the
total multiplicity $dN/dy$ and the total transverse energy $dE_T/dy$ to be equal to the experimental data. 
The charged particles multiplicity per unit of rapidity is $dN_{ch}/dy\simeq 1600 $ and the transverse energy $dE_T/dy=2100 \, GeV$ at LHC, $dN_{ch}/dy \simeq 670$ and $dE_T/dy=760 \, GeV$ at RHIC.
This leads for a radial uniform expansion to $\beta_{max}=0.37$ and $R_\perp=8.7\, fm, \tau= 4.5 \, fm/c$ at RHIC , and $\beta_{max}=0.60$ and 
$R_\perp= 10.2 \, fm, \, \tau= 7.8\, fm/c$ at LHC in quite good agreement
also with simulations in hydrodynamical or kinetic transport approaches. 
We notice that such a values correspond in one unity of rapidity  to a volume of $V\sim 1100 \hbox{ fm}^{3}$ at RHIC, while 
at LHC $V\sim 2500 \hbox{ fm}^{3}$, which means an increase of a bit more than a factor
of two in agreement with the  estimate from pion HBT interferometry \cite{Aamodt:2011mr}.

The baryon chemical potential for quarks $\mu_b=\mu_B/3$ is set to reproduce the $p/\bar p$ ratio observed experimentally which means $\mu_b=10\, MeV$ at RHIC while at LHC we have approximated $\mu_b$ to zero. 
As in Refs. \cite{Greco:2003xt,Greco:2003mm} the quark and antiquark masses considered are $m_{d,u}=m_{\bar{d},\bar{u}}=330\text{ MeV}$ for light quarks, and $m_{s}=m_{\bar{s}}=450 \,\rm MeV$ for strange quarks. 
This implies a number of quarks per unit of rapidity within the fireball at RHIC is $N_{u,d}\sim 230$, $N_{\bar{u},\bar{d}}\sim 210$ and $N_{s,\bar{s}}\sim 150$. At LHC there are $N_{u,\bar{u},d,\bar{d}}\sim 530$ and $N_{s,\bar{s}}\sim 360$.
We also notice that despite the quite different total energy at both RHIC and LHC 
the energy density  of the fireball (with radial flow energy subtracted) is
about $\epsilon \simeq 0.7\, \rm GeV/fm^3$ in good agreement with the energy density at
$T_c$ as evaluated in lattice QCD studies \cite{Borsanyi:2010cj}.

For partons at high transverse momentum, $p_T> 2.5 \, GeV$ (see previous Section), we consider the mini-jets that have undergone the jet quenching mechanism. Such a parton distribution can be obtained from pQCD calculations. 
As in Ref.\cite{Greco:2003mm} we have considered the initial $p_T$ distribution according to 
the pQCD and the thickness function of the Glauber model to go from pp collisions to
AA ones.Then we have quenched the spectra with the modeling as in Ref. \cite{Scardina:2010zz}
to reproduce the $p_T$ spectrum of pions
as observed experimentally at  $p_T \sim 8\,\rm GeV$.  
These parton spectra can be parametrized at RHIC as
\begin{equation}
\frac{dN_{jet}}{d^{2}p_{T}}=A \left( \frac{B}{B+p_{T}} \right)^{n}
\end{equation}
 which is the same used in \cite{Greco:2003xt, Greco:2003mm}
with the values given in the Table\ref{table1}.
The parametrization at LHC is
\begin{equation}
\frac{dN_{jet}}{d^{2}p_{T}}=\frac{A_{1}}{\left[ 1+ \left( \frac{p_{T}}{A_{2}} \right)^{2}\right]^{A_{3}}}+\frac{A_{4}}{\left[ 1+ \left( \frac{p_{T}}{A_{5}} \right)^{2}\right]^{A_{6}}}
\end{equation}
with the values of $A_i$ given in Table\ref{table2}
\begin{table} [ht]
\label{table1}
\begin{center}
\begin{tabular}{l c c c } 
& A$[1/GeV^{2}]$ & B[GeV] &n \\
\hline
\hline \hline
g & $3.18\cdot 10^{4}$  & 0.5 & 7.11 \\ 
\hline
$u,d$ & $9.79\cdot 10^{3}$ & 0.5 & 6.84 \\
\hline 
 $\bar{u},\bar{d}$ & $1.89\cdot 10^{4}$  & 0.5 & 7.59 \\ 
\hline 
$s$ & $6.51 \cdot 10^{3}$ & 0.5 & 7.36 \\ 
\hline
$\bar{s}$ & $8.02 \cdot 10^{3}$ & 0.5 & 7.57 \\ 
\hline \hline \hline
\end{tabular}
\end{center}
\caption{Parameters for mini-jet parton distributions at midrapidity from Au+Au at $\sqrt{s} = 200 \; AGeV$}
\end{table}

\begin{table} [ht]
\label{table2}
\begin{center}
\begin{tabular}{l c c c c c c} 
& $A_{1}$ & $A_{2}$ & $A_{3}$ & $A_{4}$ & $A_{5}$ & $A_{6}$ \\
\hline
\hline \hline
$g$ & 23.46  & 4.84 & 8.08 & 2.78 & 2.79 & 2.31 \\ 
\hline
$u,d$ & 24.68  & 5.11 & 8.01 & 0.55 & 5.65 & 2.56 \\
\hline
$\bar u, \bar d$ & 23.12  & 5.05 & 8.21 & 0.57 & 5.62 & 2.58 \\
\hline
$s$  & 24.14  & 5.11 & 8.01 & 0.55 & 5.65 & 2.56 \\
\hline
$\bar s$ & 23.12  & 5.00 & 8.31 & 0.57 & 5.62 & 2.61 \\
\hline \hline \hline
\end{tabular}
\end{center}
\caption{Parameters for mini-jet parton distributions at midrapidity from Pb-Pb at $\sqrt{s} = 2.76 \; ATeV$}
\end{table}

We note that the separation into a thermal spectrum at $p<p_0$ and a power law spectrum from hard parton process at higher $p_T$ is merely a first order approximation. In fact the parton radiating in the QGP medium create a parton shower that can lead to a coalescence process or an in-medium modification of the fragmentation function that is not accounted for in our approach. One can expect this to be particular relevant in the region of $p_T$ around 2-3 GeV at parton level that means a $p_T \approx 5 GeV$ in the hadronic spectra. Our present work based on a simplified underlying parton distribution allows indeed to spot, especially for baryons, the importance to considering a more realistic gluon radiation and splitting in the intermediate $p_T$ region. In fact in our approach we find a systematic lack of yield in such a $p_T$ region, see Section IV and V. We mention that while an approach in this direction has not yet fully settled important steps in this direction are in progress \cite{Han:2012hp,Ko:2014turic,Fries:hp2015}.

\section{Hadron spectra and ratios at RHIC}
In this section, we show results for the transverse momentum spectra
of pions, protons,antiprotons, kaons and Lambdas using the model described in 
previous sections. For the coalescence contribution, we first 
take into account the effects due to gluons in the quark-gluon plasma 
by converting them to quarks and anti-quark pairs 
with probabilities according to the flavor compositions in
the quark-gluon plasma, as assumed \cite{Biro:1994mp,Greco:2003mm}.  
For both mesons and baryons, we include not only coalescence of hard 
and soft partons as in Ref.\cite{Greco:2003mm} but also that among soft 
hard partons which are relevant for hadrons at $p_T \simeq \, 3-4\, GeV$. 
Furthermore, we include both
stable hadrons such as pion, proton (anti-proton), and kaon (anti-kaon) 
as well as the first excited resonances. With respect to \cite{Greco:2003mm} we have added
some more resonance which has allowed also to verify that the inclusion of resonances
improves the description especially at low $p_T$ but does not affect significantly
the intermediate $p_T$ and the baryon/meson ration around the peak.

\begin{figure}[ht]
\centering
\includegraphics[scale=0.32]{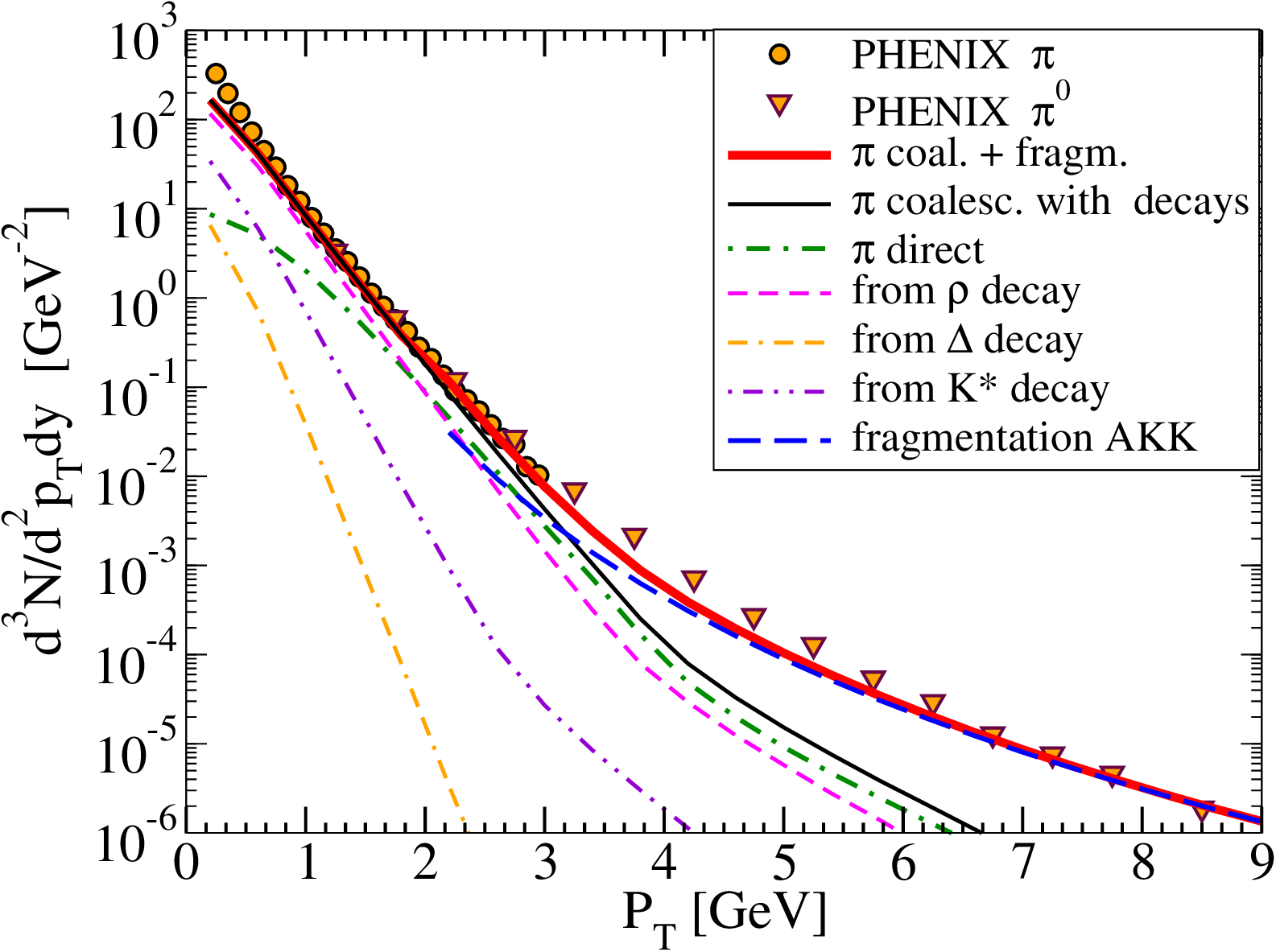}
\caption{(Color online) Pion transverse momentum spectrum at RHIC in Au+Au collisions at $\sqrt{s}=200 \,AGeV$,
$0-10\%$ centrality. Pion production from coalescence shown by thin solid line. Direct pions are shown by the dash-dotted line. Pion from resonance decay are dashed lines for $\rho$; dash double-dot line for $K^{*}$; double-dashed dot line for $\Delta$. Pion from mini-jet fragmentation are the dashed black line. Sum of both hadronization processes shown by thick solid line). Experimental data from PHENIX \cite{Adler:2003cb} \cite{Adler:2003qi}. }
\label{fig:piRHIC_decay}
\end{figure}

For the pion $\pi (I=1, J=0)$ spectrum the following resonance and their decay channels are included: $K^{*}$ (I=1, J=1/2) with $K^{*}\longrightarrow K\pi$; $\rho$ (I=1, J=1) with $\rho\longrightarrow \pi\pi$; $\omega$ (I=0, J=1) with $\omega \longrightarrow \pi\pi \pi$; $\Delta$ (I=3/2, J=3/2) with   $\Delta\longrightarrow N\pi$. 
In Fig. \ref{fig:piRHIC_decay} it is shown the predicted for the coalescence plus fragmentation 
$\pi^0$ distribution in $p_T$ for RHIC in Au+Au collisions at $\sqrt{s}=200 \,GeV$ for $(0-10\%)$ centrality; the experimental data are shown by circles \cite{Adler:2003cb} and triangles \cite{Adler:2003qi}.
We can see a general good agreement for $p_T> 1\, GeV$ that is about the region where
we expect the coalescence plus fragmentation approach to apply.
In Fig.\ref{fig:piRHIC_decay} we show much more details of the calculation. By thin solid line and
dashed line it is plot the contribution from pure coalescence and fragmentation
respectively. We can see that the contribution of both mechanism becomes about similar for
$p_T \simeq 3.5 \, \rm GeV$ and we will see that for the baryon instead the coalescence will dominate
in a large $p_T$ range. We can also see the contribution for the decay into pions coming from resonances, which shows that contribution from $\rho \rightarrow \pi \pi$, dashed line, dominates up to about $p_T \sim 3 \,\rm GeV$ which is about the region where anyway the fragmentation is starting
to take over. The contribution from $K^*$ (dashed double-dotted line) and $\Delta$ (double-dashed dot line) are instead quite less relevant and only
contribute to some little improvement of the description at very low $p_T$.
Of course for the pions it is know more or less all the hadrons contribute to the feed-down, see also Ref.\cite{Greco:2004ex}, but in the region we are interested in the resonances included are sufficient to have a good description
of the pion spectra at $p_T\gtrsim 1\, \rm GeV$, see Fig.\ref{fig:piRHIC_decay}.

\begin{figure}[ht]
\centering
\includegraphics[scale=0.32]{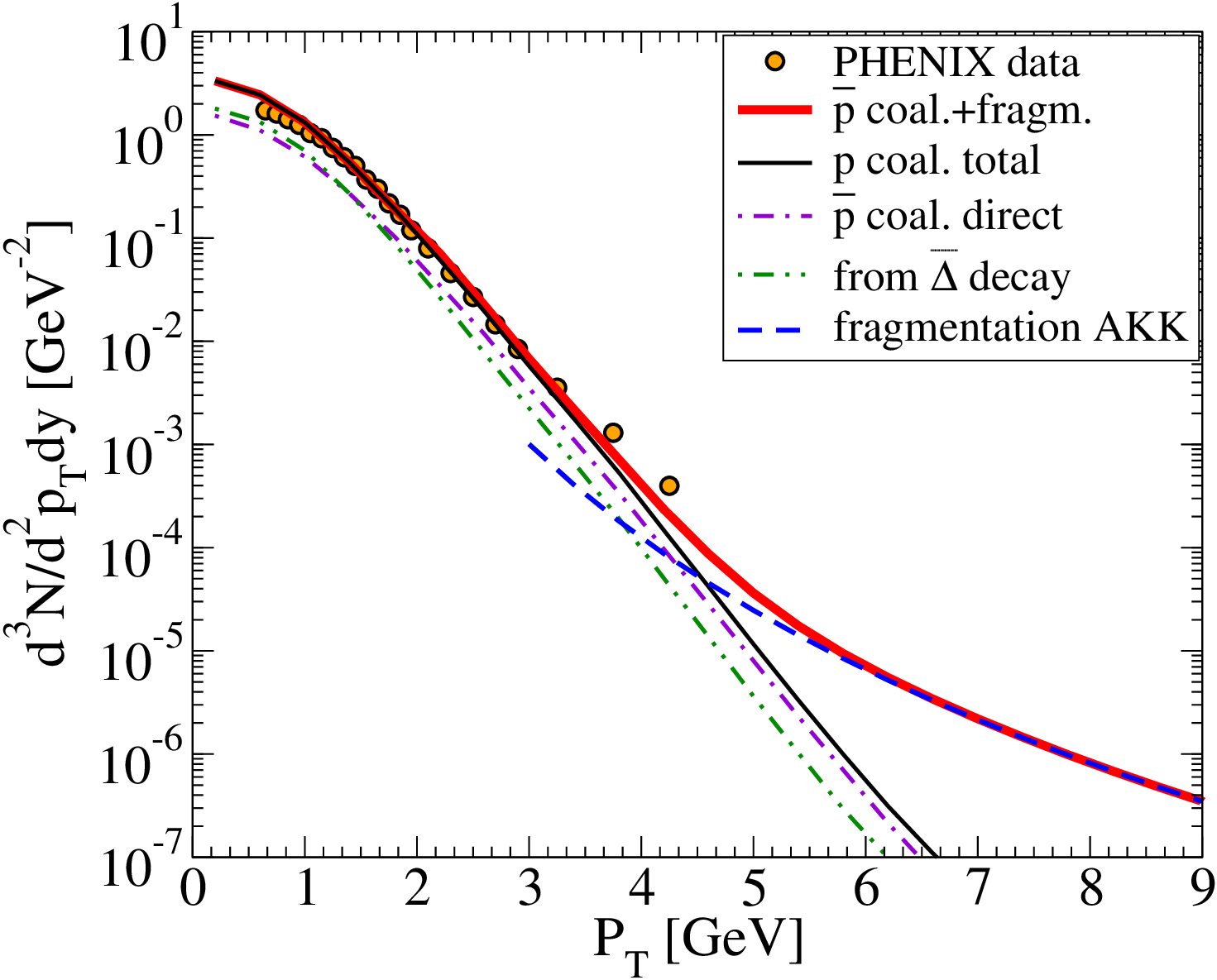}
\caption{(Color online) Antiproton transverse momentum spectrum at RHIC in Au+Au collisions at $\sqrt{s}=200 \;AGeV$,
$0-10\%$ centrality. Antiproton production from coalescence thin solid line. Direct antiproton are shown by the violet dash-dotted line; from $\Delta$ decay are the dash double-dotted line. Antiproton from mini-jet fragmentation are the dashed line. Sum of both hadronization processes shown by thick solid line. Experimental data from PHENIX \cite{Adler:2003cb}. }
\label{fig:aprotRHIC_decay}
\end{figure}
\begin{figure}[ht]
\centering
\includegraphics[scale=0.32]{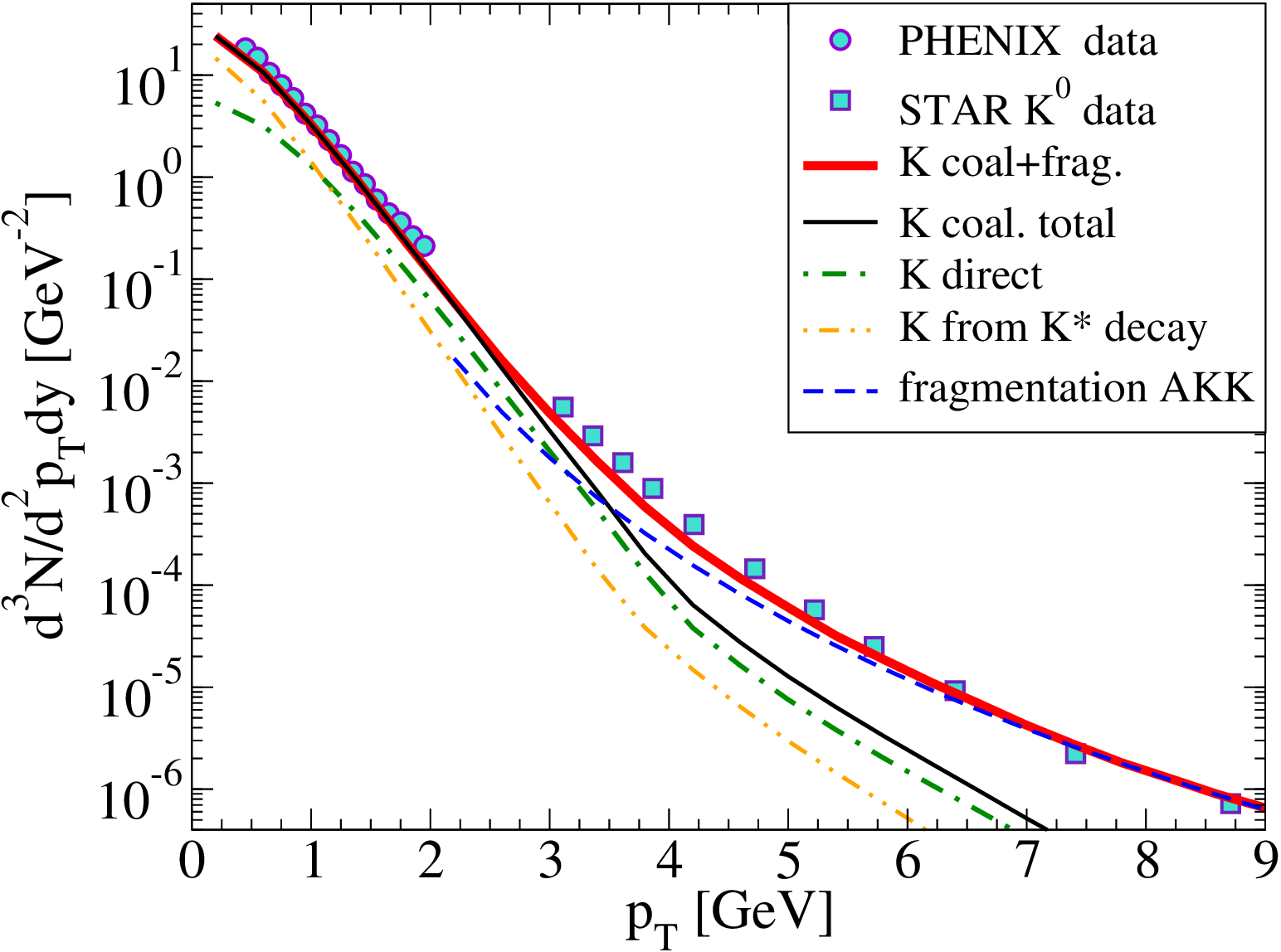}
\caption{(Color online) Kaon transverse momentum spectrum at RHIC in Au+Au collisions at $\sqrt{s}=200 \;AGeV$,
$0-10\%$ centrality. Kaon production from coalescence is the thin solid. Direct kaons are shown by the dash double dotted line. Kaons from $K^{*}$ decay are the dash-dotted line. Kaon from mini-jet fragmentation is the dashed line. Sum of both hadronization processes is shown by thick solid line. Experimental data from PHENIX \cite{Adler:2003cb} are shown by the circles, STAR \cite{Agakishiev:2011dc} data are shown by the squares. }
\label{fig:kRHIC_decay}
\end{figure}

For the  proton (anti-proton) (I=1/2, J=1/2) we have included the
$\Delta$ (I=3/2, J=3/2) with its decay, $\Delta\longrightarrow N\pi$, but we have also checked 
that adding the next exited state $N(1440)$ (I=1/2, J=1/2) production and decay leads to negligible contribution due to the large suppression factor but also to the
fact that about $40\%$ of the decay goes through $N(1440)  \longrightarrow N\pi\pi$ which
means that the proton spectrum coming from $N(1440)$ is quite steep and gives some contribution
only at quite low $p_T\lesssim 0.5 \, \rm GeV$.
In Fig. \ref{fig:aprotRHIC_decay} we show the anti-proton transverse momentum spectrum at RHIC
including coalescence and fragmentation by thick solid line together with the available experimental
data (circles) from Refs. \cite{Adler:2003cb}. Again the description appears to be quite good; we show also the
relative contribution from coalescence and fragmentation by thin solid line and by dashed line respectively. We notice that for anti-protons the two mechanism become comparable at $p_T \simeq 5 \, \rm GeV$ which means that the coalescence contribution is more important for protons with respect to pions.

In Fig. \ref{fig:kRHIC_decay} we show also the $p_T$ distribution for the $K^{\pm}$ (I=0, J=1/2)
for which we have included the $K^{*}(892)$ (I=1, J=1/2) contribution from the decay of
$K^{*}\longrightarrow k\pi$. The impact of the next excited state $K_1(1270)$ is again negligible especially at $p_T\gtrsim 1\, \rm GeV$. We can see that also for Kaons the agreement with experimental data from PHENIX at low $p_T$ \cite{Adler:2003cb}, open circles, and STAR \cite{Agakishiev:2011dc}, squares, is fairly good in all the range of $p_T$.
By dash double dotted line in Fig.\ref{fig:kRHIC_decay}, we see that at low $p_T$ the contribution from $K^*$ decay becomes important and contributes to have the correct slope of the spectrum as measured experimentally. One can notice as in the case of 
pions that there is some lack of yield at $p_T \simeq \, 4 \, \rm GeV$ where the fragmentation
is starting to be dominant. We anticipate that such a systematic is observed also at LHC and 
from the ratio baryon/meson we will see that it is even more marked for baryons and in particular
for Lambda's. 
\begin{figure}[ht]
\centering
\includegraphics[scale=0.32]{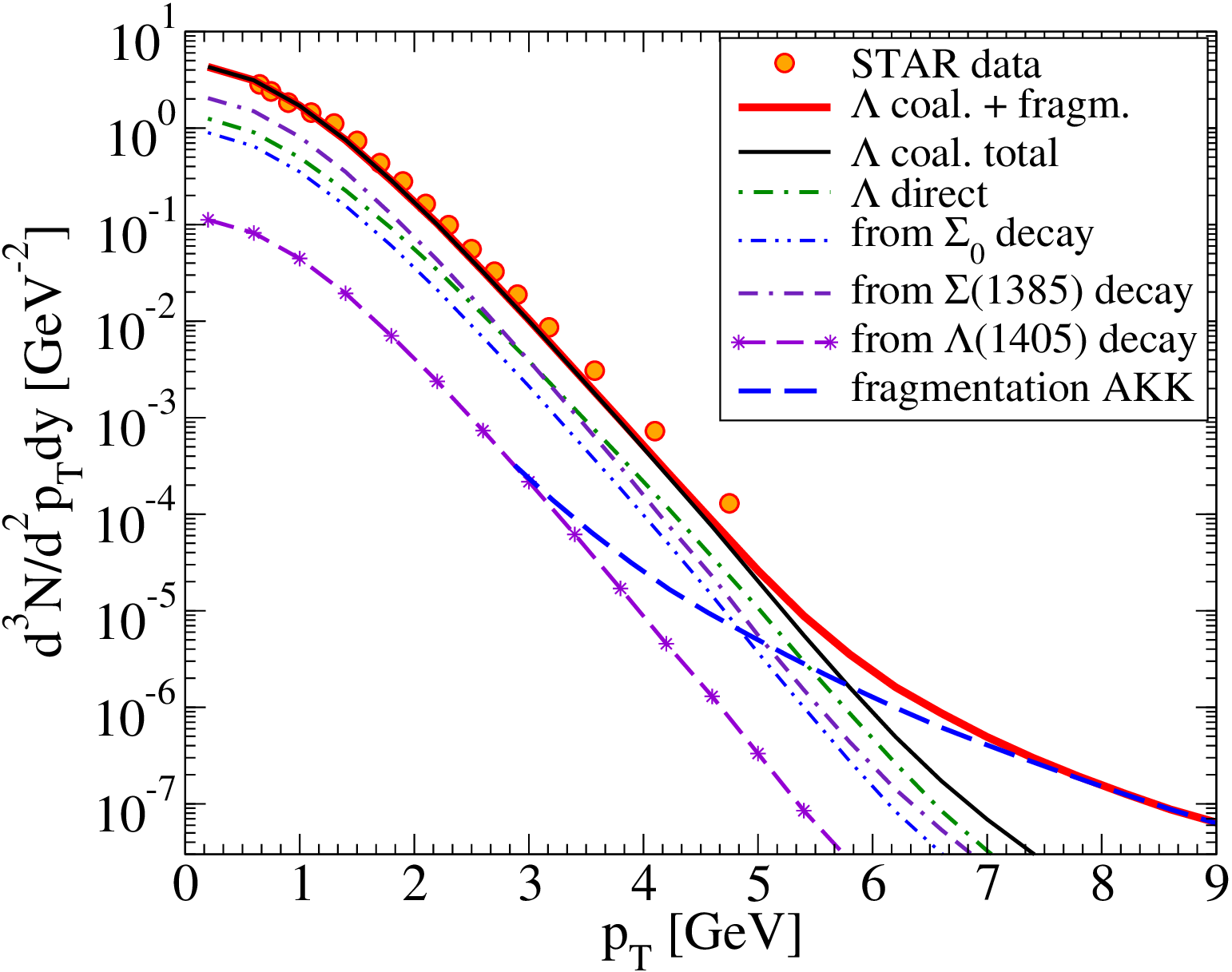}
\caption{(Color online) $\Lambda$ transverse momentum spectrum at RHIC in Au+Au collisions at $\sqrt{s}=200 \;GeV$, $0-10\%$ centrality. $\Lambda$ production as sum of both hadronization processes is shown by thick solid line; from coalescence only by thin solid line. Direct $\Lambda$\textit{s} are shown by the dash-dotted line. $\Lambda$ from resonance decay are: $\Sigma^{0}$ (dashed double dotted line), $\Sigma(1385)$ (double dashed dotted line), $\Lambda(1405)$ (dashed line with $*$ symbols). $\Lambda$ from mini-jet fragmentation is the dashed line. Experimental data from STAR \cite{Adams:2006ke}. }
\label{fig:LamRHIC_decay}

\end{figure}
The $p_T$ distribution for $\Lambda(1116)$ (I=0, J=1/2) at RHIC energy in Au+Au collisions at $\sqrt{s}=200 \;GeV$
for central collisions is shown in Fig. \ref{fig:LamRHIC_decay} by thick solid line along with
the experimental shown by circles \cite{Adams:2006ke}. For the $\Lambda$ there are indeed several
hadronic states that have a significant contribution and we have included the following closest resonances with their decay channels and the pertinent branching ratios (B.R.): $\Sigma_{0}(1193)$ (I=1, J=1/2), $\Sigma_{0}\longrightarrow\Lambda \gamma$; $\Lambda(1405)$ (I=0, J=1/2), $\Lambda(1405)\longrightarrow\Sigma \pi$; $\Sigma(1385)$ (I=1, J=3/2) with the two main decay channels: $\Sigma(1385)\longrightarrow\Lambda \pi$ with $B.R.=88\%$ and $\Sigma(1385)\longrightarrow\Sigma \pi$ with $B.R.=11.7\%$. In Fig. \ref{fig:LamRHIC_decay} one can also find the different contribution of these
channel, see also the caption. 
Also for the Lambda the coalescence plus fragmentation model appear to be able to correctly described 
the experimental data  in a wide range of $p_T$. 
We also find similarly to the anti-proton that the contribution
from independent fragmentation according the AKK parametrization becomes dominant at 
$p_T \gtrsim 6 \,\rm GeV$. 

The coalescence mechanism has had the merit to naturally predict a baryon/meson enhancement
at intermediate transverse momentum, especially in the region $p_T\simeq 2-4 \, \rm GeV$ where
the $p/\pi^+$, $\bar p/\pi^-$, $\Lambda/2\,K^0_s$ reaches a value of the order of unity which
is a strong systematic enhancement with respect to the one 
observed in pp collisions \cite{Fries:2008hs}. 

\begin{figure}[ht]
\centering
\includegraphics[scale=0.32]{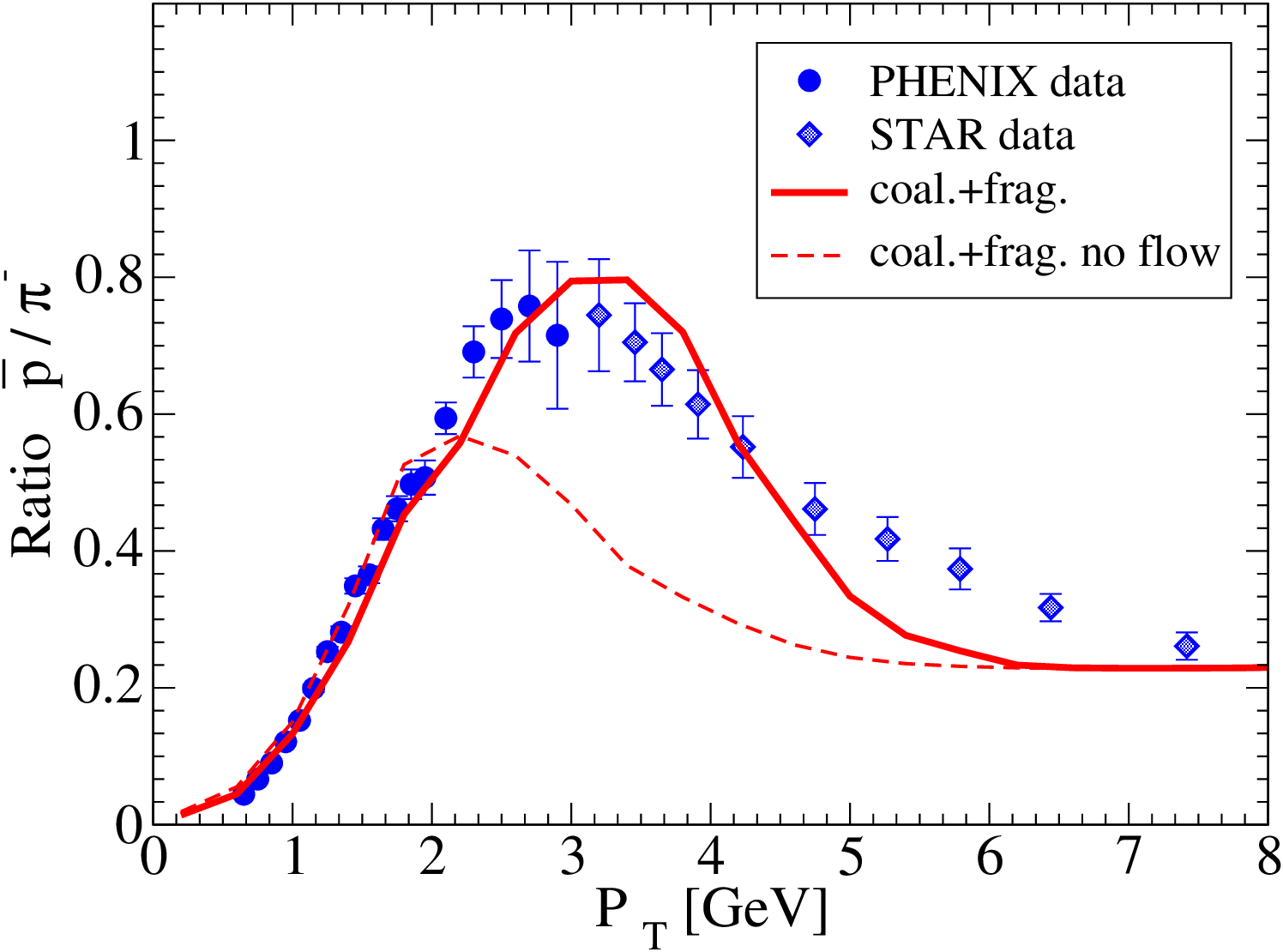}
\caption{(Color online) Anti-Proton to positive pion ratio at RHIC from Au+Au collisions at $\sqrt{s}=200\;GeV$.  The model prediction is the thick solid line; by dashed line the result switching off the radial flow of the QGP bulk matter. PHENIX \cite{Adler:2003cb} data are shown by circles, STAR \cite{Agakishiev:2011dc} data by triangles. }
\label{fig:ratioppi_rhic}
\end{figure}

\begin{figure}[ht]
\centering
\includegraphics[scale=0.32]{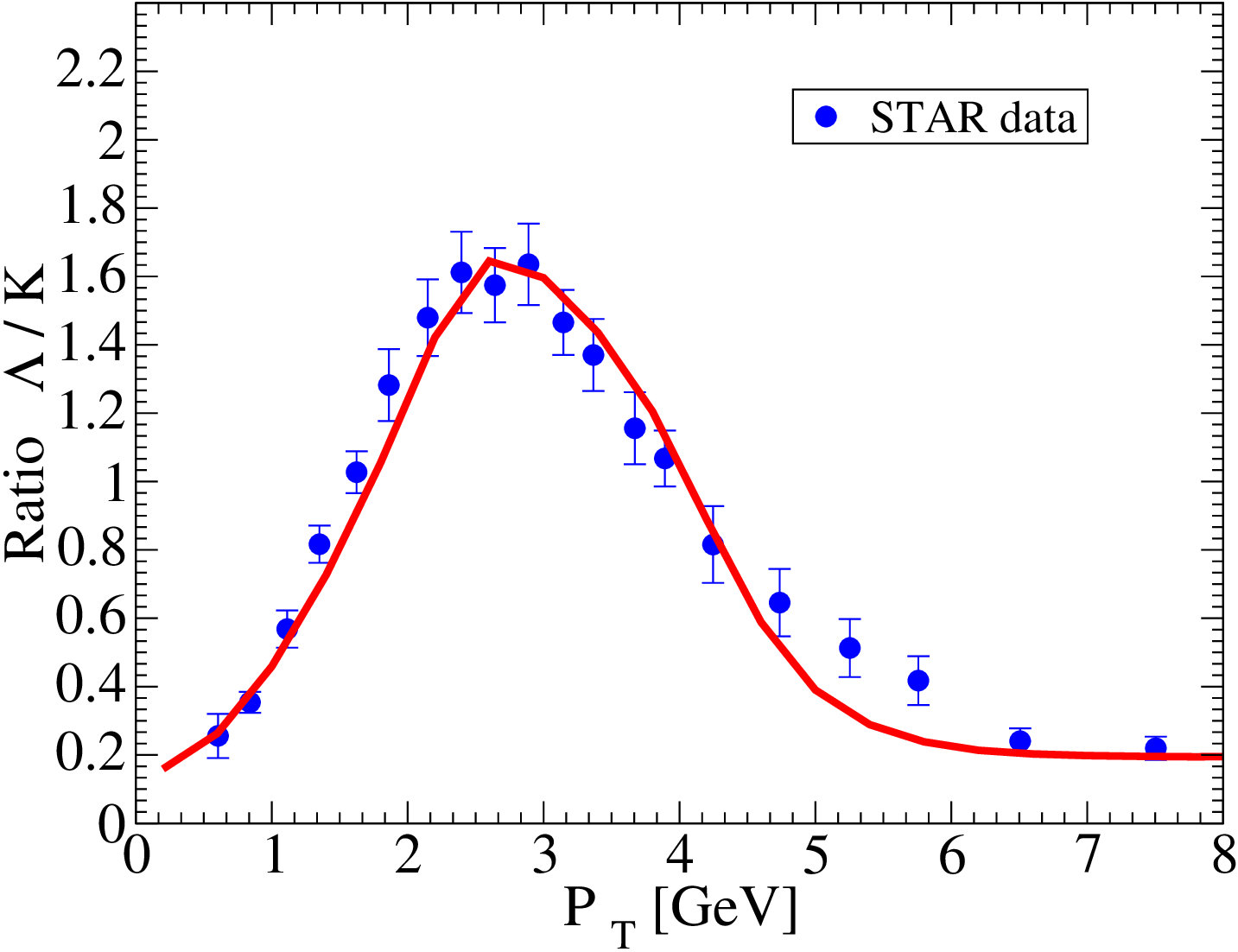}
\caption{(Color online) Lambda to kaon ratio at RHIC from Au+Au collisions at $\sqrt{s}=200\;\rm AGeV$.  The model prediction is the solid line. STAR data by circles \cite{Agakishiev:2011ar}. }
\label{fig:ratiolamk_rhic}
\end{figure}

We therefore show the $\bar p/\pi^-$, $\Lambda/K^0_s$ in the Fig.s \ref{fig:ratioppi_rhic}
and \ref{fig:ratiolamk_rhic}. We can see that the ratio is quite well predicted from its
rise at low $p_T$ up to the peak region and then the falling-down behavior. However in both cases
it is clear that in the region of $p_T\simeq 5-7\,\rm GeV$ there is a lack of baryon yield.
This is a feature that could not be observed when the coalescence+ fragmentation model was
applied a decade ago to hadronization at RHIC because there were no data available
for proton (anti-proton) at $p_T \gtrsim 4 \,\rm GeV$, nonetheless it appears systematically, 
we will see it also at LHC energy, so we postpone some further comment about it at the
end of the next Section.

\section{Hadron spectra and ratios at LHC}

We present in this Section the results from the coalescence plus fragmentation and the comparison 
with the experimental data for Pb+Pb collisions at $\sqrt{s}=2.76\; \mbox{ATeV}$ for $0-10\%$
collisions. We notice that the results are obtained without any change or addition of parameters
with respect to the one at RHIC in the previous Section. Indeed a main aim of the present
paper is to show that the evolution of the baryon/meson pattern from RHIC to LHC can be 
correctly and self-consistently predicted. The only change with respect to RHIC is 
in the radial flow and volume of the hadronizing fireball that as described above 
is self-consistently constrained but the total transverse energy and the multiplicity that
at LHC $\sqrt{s}=2.76\,\rm ATeV$is about a factor 2.4 and 2.7 larger with respect to RHIC $\sqrt{s}=200\,\rm AGeV$, see Section III-a.

\begin{figure}[ht]
\centering
\includegraphics[scale=0.32]{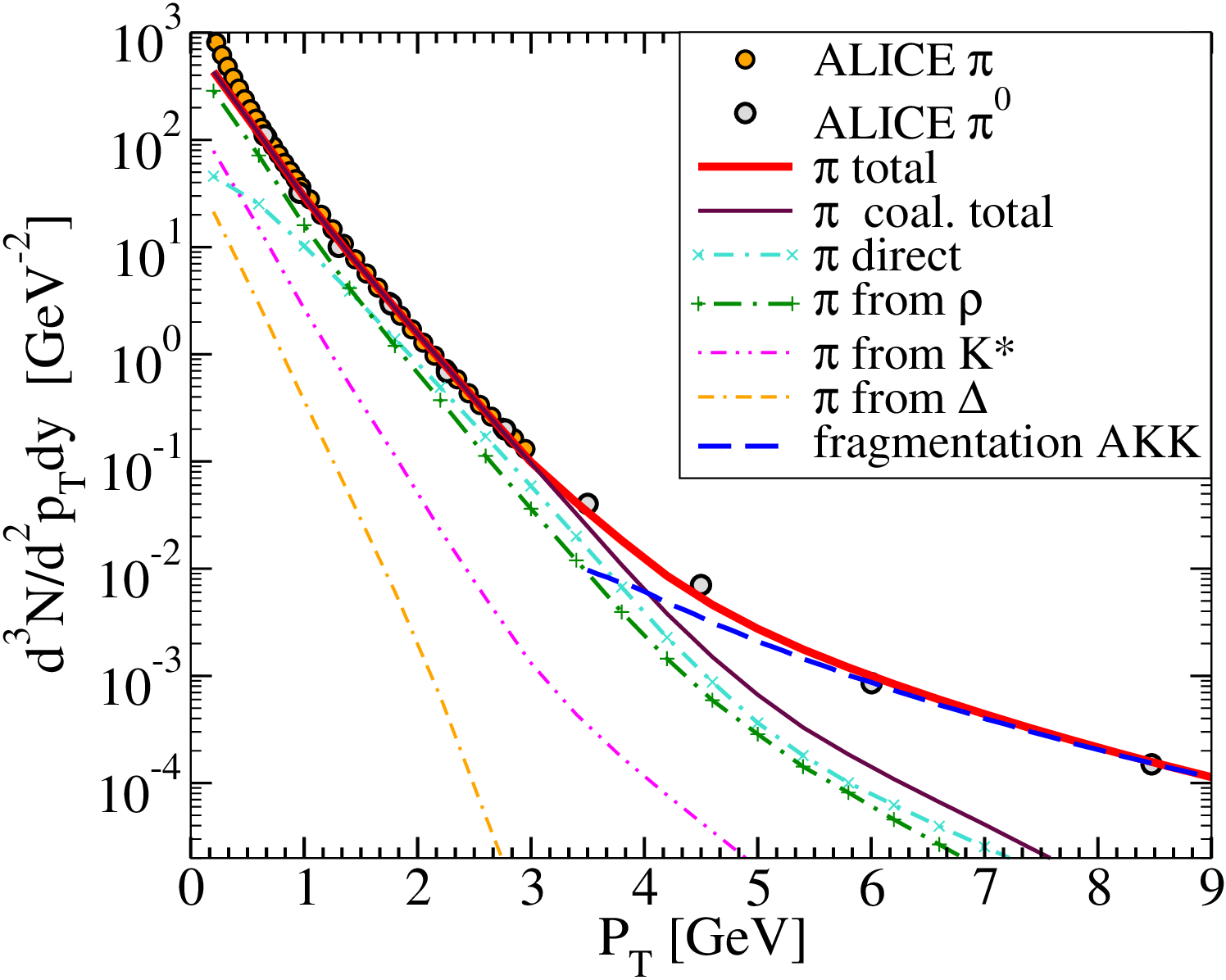}
\caption{(Color online) Pion transverse momentum spectrum from Pb+Pb collisions at $\sqrt{s}=2.76\; \mbox{TeV}$,
$0-10\%$ centrality. The solid curve includes contributions from coalescence process. The thick solid line is the sum of coalescence and fragmentation contributions. Direct pions are shown by the dash-dotted line with X symbols. Pion from resonance decay are the dash-dotted lines with "+" symbol for $\rho$ , dashed double-dotted lines for $K^{*}$, double-dashed dotted line for$\Delta$. Pions from mini-jet fragmentation are the dashed line. The circles are the experimental data from ALICE experiment \cite{Abelev:2012wca} \cite{Abelev:2014ypa}. }
\label{fig:pionLHC}
\end{figure}

In Fig.\ref{fig:pionLHC} we show the total (coalescence + fragmentation) pion spectrum by thick solid line which is a quite
good agreement with the experimental data on $\pi^0$ from ALICE Collaboration in all the 
$p_T$ range, except some lack of yield at $p_T\lesssim 0.5\,\rm GeV$ due to absence of all the resonance
decays feed-down. By thin solid and dashed line we show the contribution from coalescence and
fragmentation respectively. We notice that the two yields cross at $p_T \simeq 4 \, \rm GeV$ which is about a shift of about 1 GeV with respect to RHIC, see Fig. \ref{fig:piRHIC_decay} such a shift is due to the larger collective flow present at LHC that shifts to larger $p_T$ the hadrons from coalescence .
The very good agreement of the $p_T$ distribution at LHC already shows that the model
is able to correctly predict the evolution of the absolute yield and especially its $p_T$
shape correctly, in fact no parameter of the coalescence process, essentially the Wigner wave function
width $\Delta_p$ of the hadrons, has been modified with respect to those used in the previous Section for RHIC.

\begin{figure}[ht]
\centering
\includegraphics[scale=0.32]{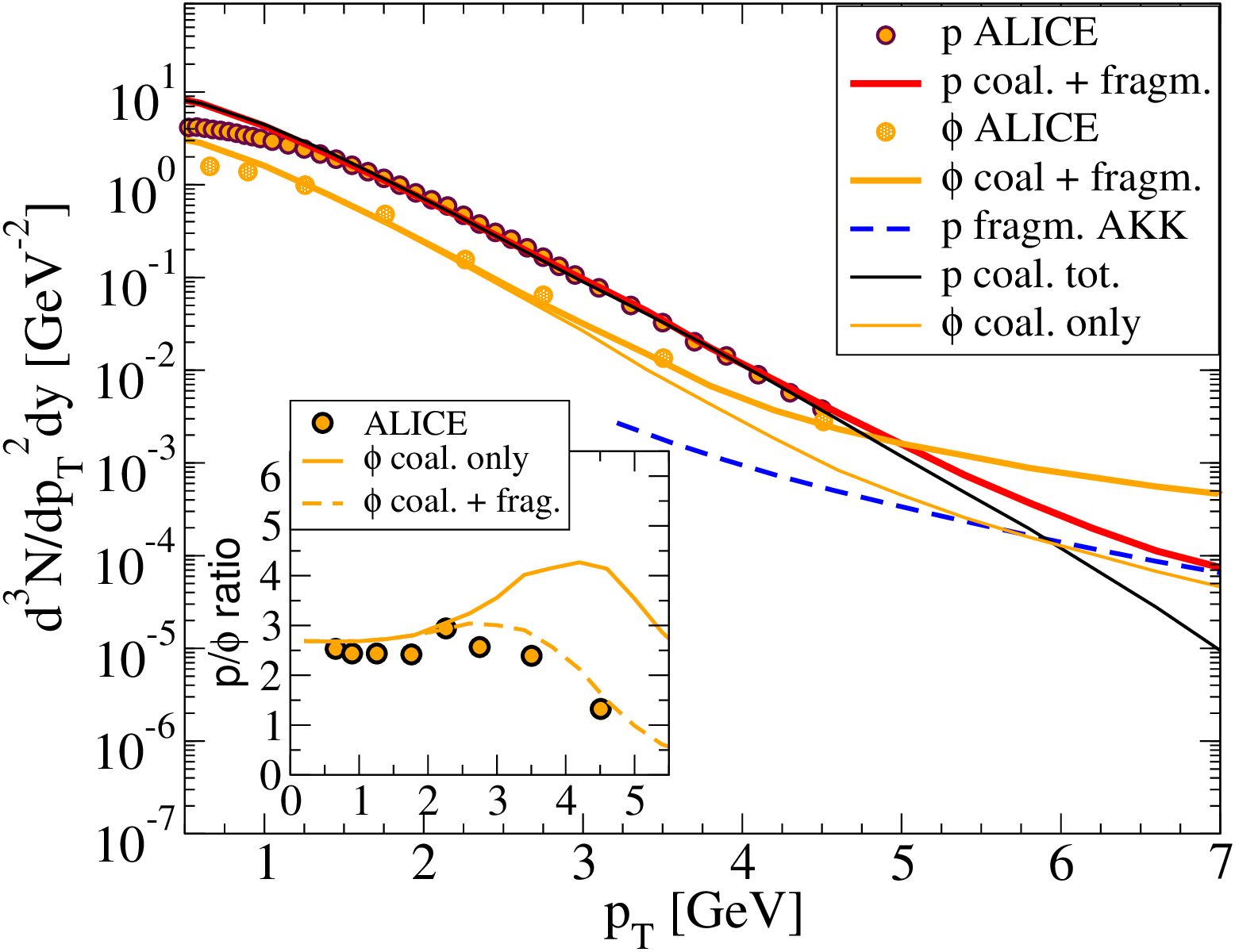}
\caption{(Color online) Proton and $\phi$ transverse momentum spectrum from Pb+Pb collisions at $\sqrt{s}=2.76\;\mbox{ATeV}$,
$0-10\%$ centrality. The solid thick line is the sum of coalescence and fragmentation process, for $\phi$
see the text. The solid black thin curve shows the contribution from coalescence. Protons from mini-jet fragmentation are the dashed line. The orange solid thick line is the total $\phi$ spectrum and the light thin line is the contribution from only coalescence, see text.  The circles are the experimental data from ALICE experiment \cite{Abelev:2012wca,Abelev:2014uua}. In the inset the $p/\phi$ ratio is shown.}
\label{fig:proton}
\end{figure}

\begin{figure}[ht]
\centering
\includegraphics[scale=0.32]{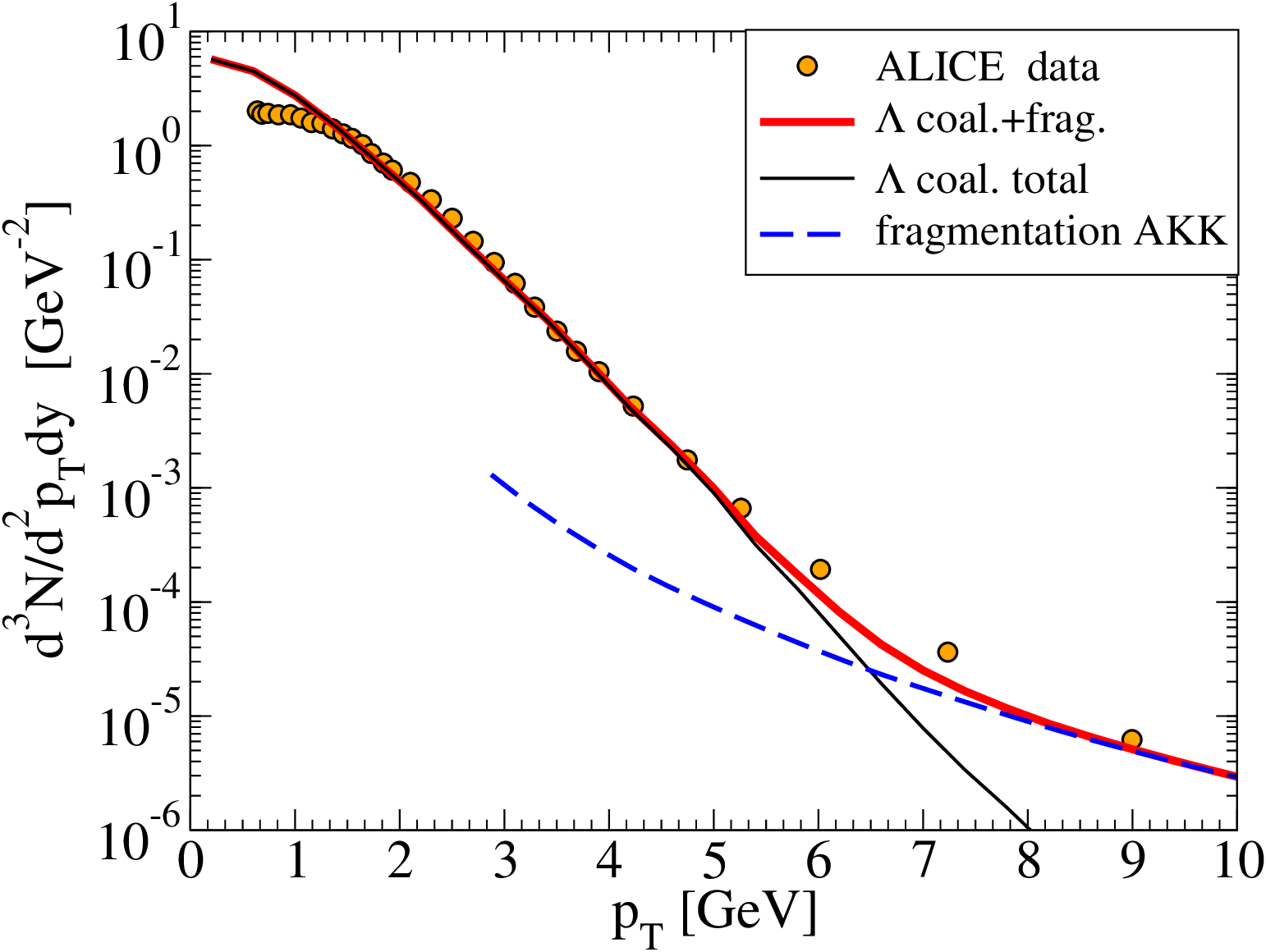}
\caption{(Color online) $\Lambda$ transverse momentum spectrum from Pb+Pb collisions at $\sqrt{s}=2.76\; \mbox{ATeV}$,
$0-10\%$ collisions. The thin solid curve includes all contributions from coalescence process.$\Lambda$ from mini-jet fragmentation is the dashed line.  The sum of coalescence and fragmentation contributions is shown by the thick solid line. The circles are the experimental data from ALICE experiment \cite{Abelev:2013xaa} \cite{Chinellato:2012jj}. }
\label{fig:lambda}
\end{figure}

In Fig.\ref{fig:proton} it is shown the proton spectrum at LHC by thick solid line and compared
to the experimental data. The agreement also in this case is very good for $p_T > 1\,\rm GeV$ 
up to 5 GeV that is the maximum value with available data.
At very low $p_T$ as said in Section II we should not expect the approach to really apply. Still
we can notice that the coalescence over predict the yield. The effect was partially present also
at RHIC, see Fig. \ref{fig:aprotRHIC_decay} and it is found also for the $\Lambda$ spectrum 
in Fig.\ref{fig:lambda}. It however appears in the quite low $p_T$ region that is
were as said in the introduction the present implementation of coalescence 
is not expected to work.

In Fig.\ref{fig:proton} we can also see that the yield of the fragmentation process becomes 
comparable to the one from coalescence at a $p_T \simeq 6\,\rm GeV$ which is about a $50\%$
larger with respect to the pions and also a shift of about 1.5 GeV with respect to RHIC.
This is what one would expect due to the larger flow at LHC and the fact that baryon
are more affected by it.

\begin{figure}[ht]
\centering
\includegraphics[scale=0.32]{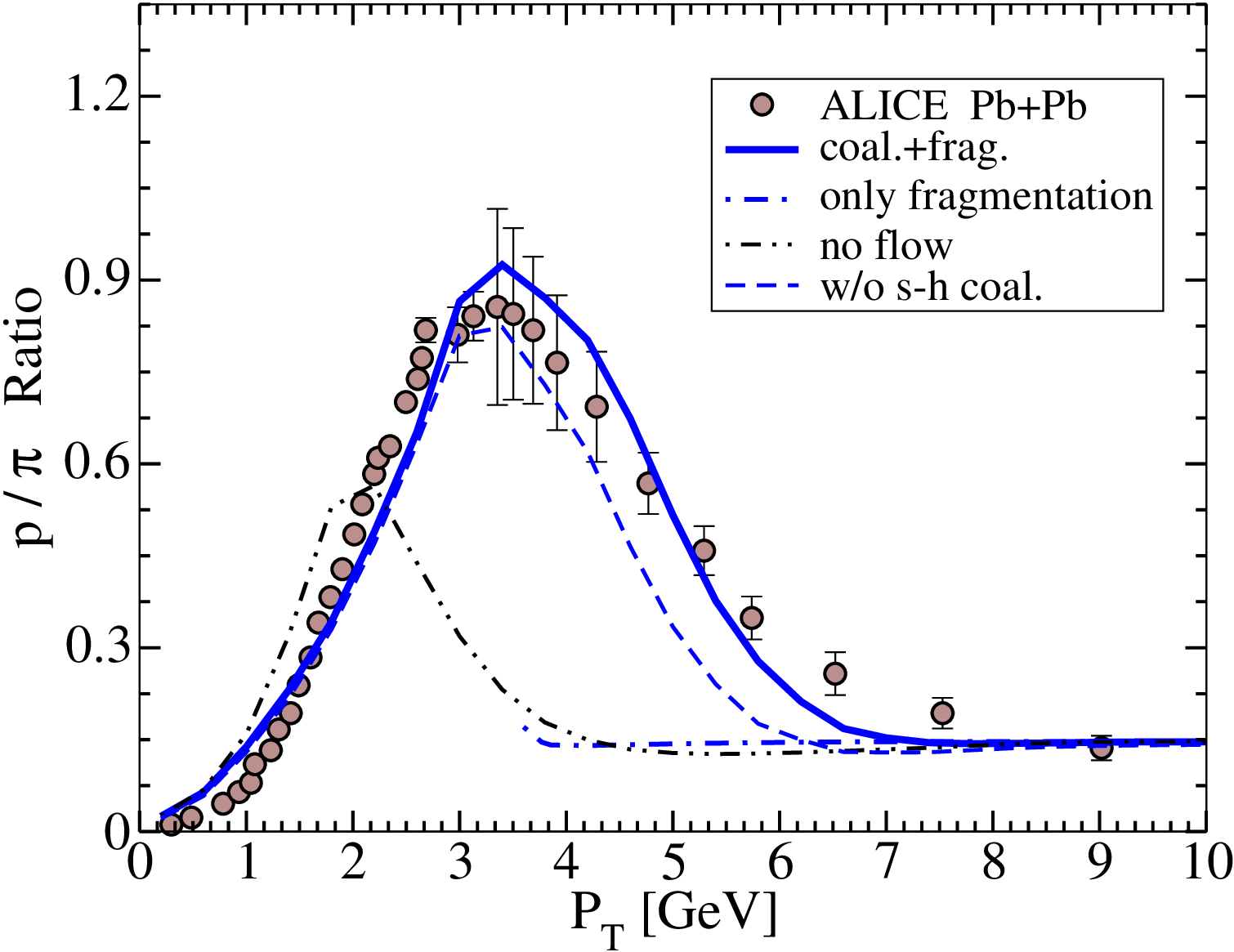}
\caption{(Color online) Proton to pion ratio in Pb+Pb collisions at $\sqrt{s}=2.76 \;\rm ATeV$. The solid line is the prediction of our model. The empty circles are data from ALICE experiment in collision Pb+Pb at 0-5\% centrality \cite{Abelev:2014laa}.}
\label{fig:ratioppi}
\end{figure}

\begin{figure}[ht]
\centering
\includegraphics[scale=0.32]{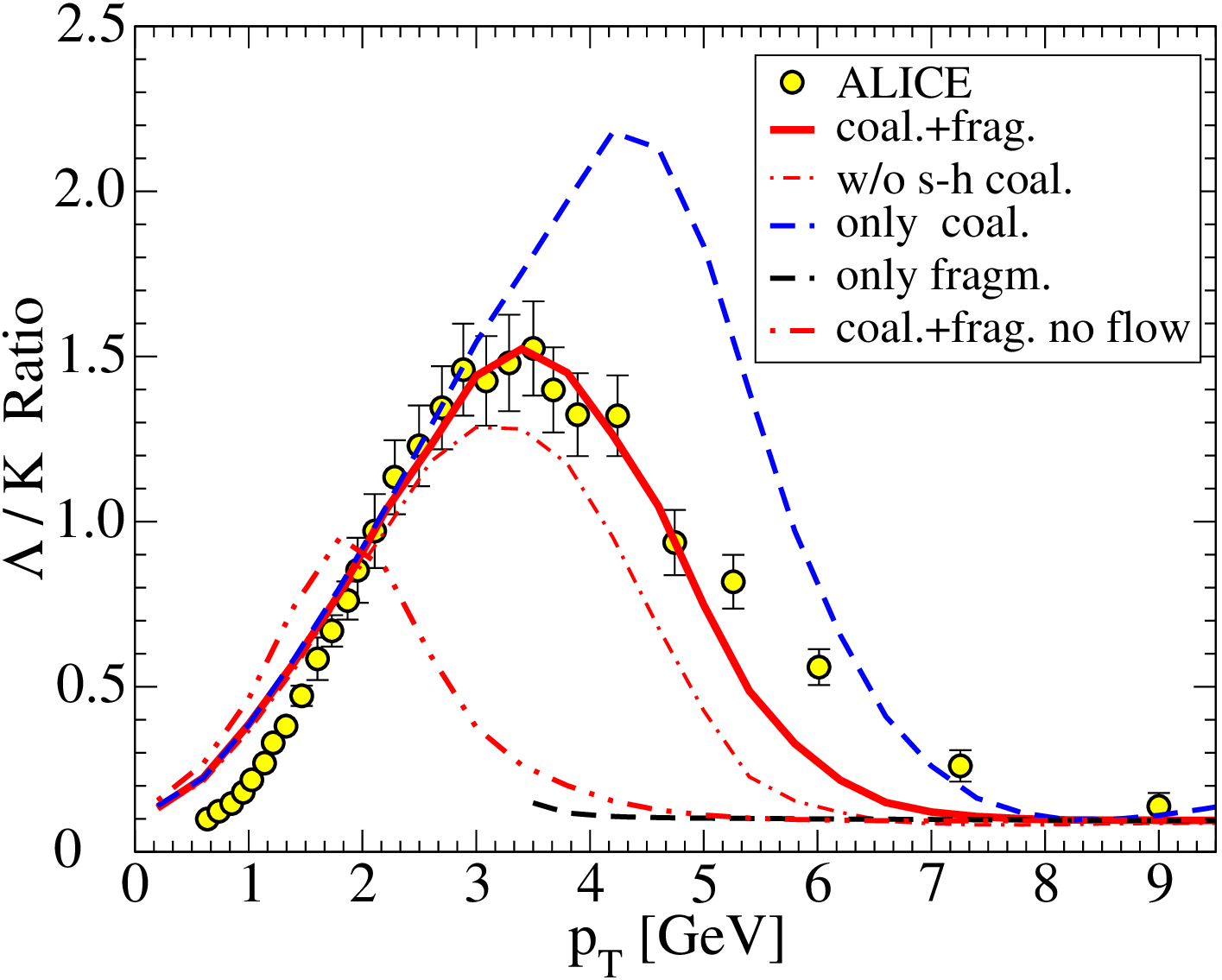}
\caption{(Color online) Lambda to kaon ratio in Pb+Pb collisions at $\sqrt{s}=2.7 \, ATeV$. The solid line is the prediction of our model. The circles are data from ALICE experiment in collision Pb+Pb at 0-5\% centrality. \cite{Abelev:2013xaa}}
\label{fig:ratiolamk}
\end{figure}

In Fig. \ref{fig:proton} we show also the $\phi$ meson spectra obtained rescaling the width parameter 
$\Delta^{meson}$, see Section II.a, according to harmonic oscillator width $\sqrt{m_q/m_s}$. It is often discussed
if the $p_T-$spectra of $\phi$ meson would have a slope close to the one of the proton
like in a hydro picture or would behave like other mesons being formed by two quarks.
We briefly mention that indeed also in a coalescence process one can/should expect that
there is a radial flow mass effect like in a hydro picture. In fact for a proton there is a combination
of 3 quark flowing with a mass of about 330 MeV  while for a $\phi$ meson there are 2 quark
flowing with a mass of about 550 MeV. The difference between this two cases is of course
only marginal, in fact we can see in Fig. \ref{fig:proton} that at low $p_T$
the slope of $\phi$, orange thin solid
line, is quite similar to the one of the proton. In the inset we show more in detail the $p/\phi$
ratio including only coalescence for $\phi$ by solid line. We can see that at $p_T \lesssim 2 \,\rm GeV$
the ratio is nearly flat. At higher $p_T$ there is a peak at about 4 GeV which signals that the slope
of the $\phi$ is stiffer, on the other hand as we see for the other mesons at such a $p_T$ there
is a significant contribution from fragmentation. Unfortunately there are no $\phi$ AKK (or KKP)
fragmentation function and it is not possible to perform a solid prediction at higher $p_T$.
However considering that the prediction for the $\phi$ in a coalescence plus fragmentation approach
are particularly awaited, we show in the inset by dashed line, and by thick solid line in the main panel,
what would be the $p_T$ distribution of $\phi$ if one add a fragmentation corresponding to the
same fragmentation over coalescence ratio as for the $K^+$ meson. We can see that as for the 
other ratios we have a quite good agreement with the data by ALICE shown by
circles \cite{Abelev:2014uua}, see also Fig. \ref{fig:ratioppi} and \ref{fig:ratiolamk}.

\begin{figure}[ht]
\centering
\includegraphics[scale=0.32]{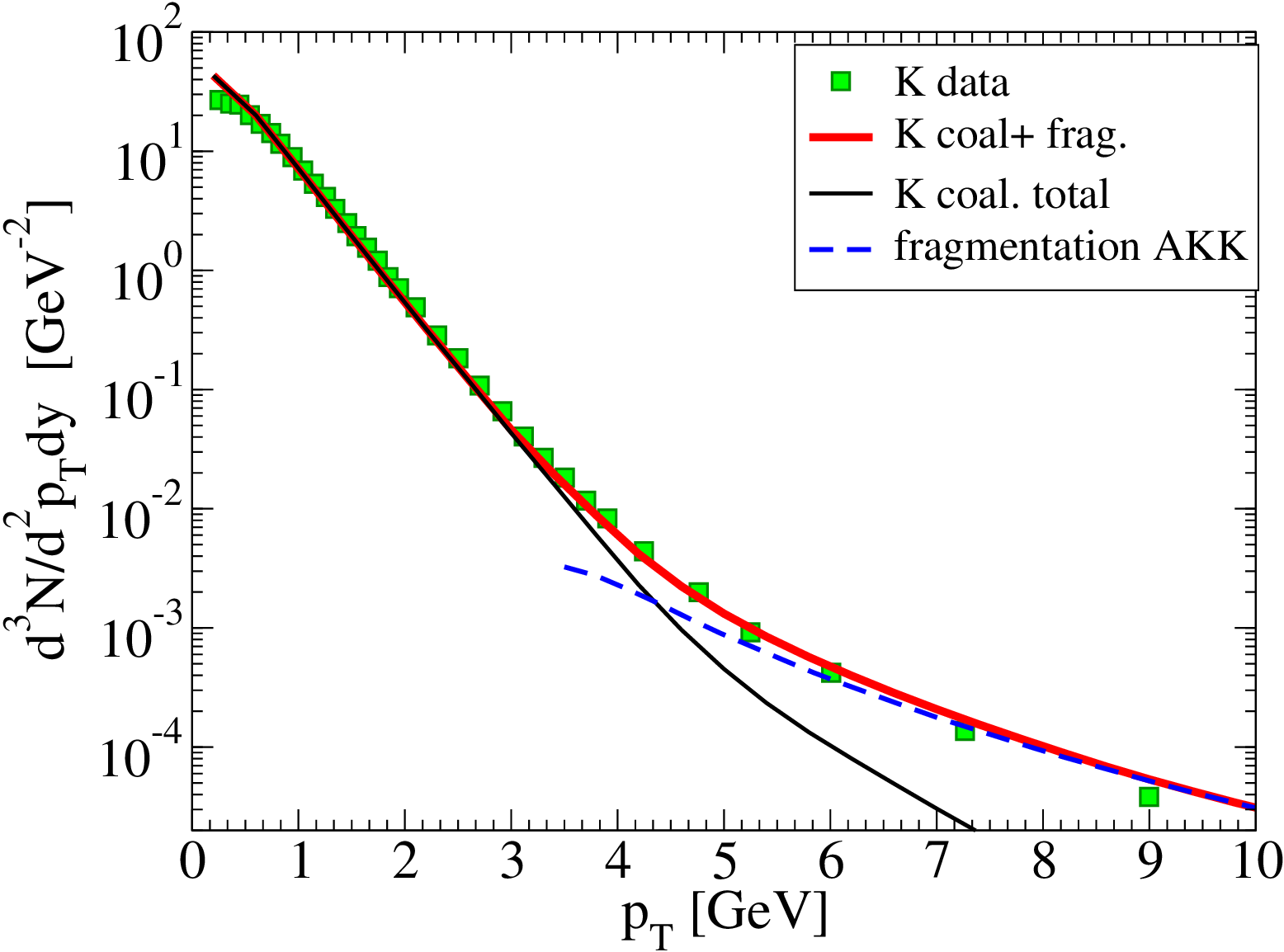}
\caption{(Color online) Kaon transverse momentum spectrum from Pb+Pb collisions at $\sqrt{s}=2.7\; \mbox{ATeV}$,
$0-10\%$ centrality. The thin solid curve includes contributions from coalescence process. The thick solid line is the sum of coalescence and fragmentation contributions. Kaons from mini-jet fragmentation are the  dashed line. The squares are the experimental data from ALICE experiment  \cite{Abelev:2012wca,Abelev:2013xaa,Chinellato:2012jj}. }
\label{fig:kaon}
\end{figure}

In Fig.\ref{fig:kaon} the $p_T$ distribution for $K^+$ is shown by thick solid line and again
one can see the good agreement with the experimental data \cite{Chinellato:2012jj} in the entire
range of $p_T$. We can notice that at RHIC for both pions and kaons there was some lack
of yield in the region where the fragmentation takes over, while at LHC energy for both
cases the agreement appears quite better. This can be expected because the independent fragmentation
function picture should be better constrained at energies of the order of TeV.

In Fig.\ref{fig:lambda} the experimental data for the transverse momentum spectrum of $\Lambda$ is
shown by circles together with the results from the coalescence plus fragmentation shown
by thick solid line. The different contribution from excited state have been calculated
and have a similar relative contribution as at RHIC, see Fig.\ref{fig:LamRHIC_decay} We can see also for this case the good agreement for $p_T> 1 \, \rm GeV$,
but while for $p$ and $\bar p$ the data are available only up to 4-5 GeV, in this case we have the availability of data up to 9 GeV and this allows us to see that in the $p_T$ region where
the fragmentation starts to dominate, $p_T \simeq \, 6-7\rm\, GeV$ there is some lack of yield.
This something that we could only marginally spot at RHIC energy mainly through the
$\Lambda/K$ ratio, see Fig. \ref{fig:ratiolamk_rhic}. 
At both RHIC and LHC such a lack of yield appears where coalescence becomes less important
therefore one can say that it seems that the spectrum from AKK fragmentation function
appears too flat. This may very well be because the fragmentation function for baryons in general
and in particular for $\Lambda$ are known to be not very well constrained. 
On the other hand we notice that the fragmentation contribution has been calculated
for all hadrons considered  with the same $Q^2=(p_{had}/2z)^2$ and this gives a global good description
of the spectra for $\pi,K$ at $p_T> 5\,\rm GeV$ and for $p,\bar p, \Lambda$ for $p_T>8\,\rm GeV$.
However we mention that a similar problem has been recently pointed-out also in pp collisions
in Refs. \cite{d'Enterria:2013vba}; also it seems that the KKP fragmentation can supply a better description even if still not satisfying. We also notice that in pp collisions the fragmentation function gives an excess of yield increasing with $p_T$ while in AA such an effect is
hidden by the large jet quenching that is tuned to reproduce the pion spectra at 
$p_T \simeq \, 8\, \rm GeV$. On the other it still appears as a too flat spectrum
at $p_T \sim 3-8 \,\rm GeV$. 
In AA collisions
it is likely that studies of in-medium fragmentation function can find a solution \cite{Werner:2012sv,Han:2012hp}
or it could be that coalescence contribution should extend to large $p_T$ 
with respect to the present modeling having simple spheres in momentum space as Wigner function and no dynamical role of the interaction that could lead to an extension of the coalescence
to pair with larger relative momentum.

We mention that recently it has been developed a process that within the coalescence
plus fragmentation approach could be quite important in solving this issue \cite{Han:2012hp}.
The idea is to describe the in-medium fragmentation as a quark recombination of shower partons 
taking into account also the gluon splitting into quark pairs that recombine.
Such a mechanism seems to provide a large contribution for baryons in the region
of $p_T \sim 6\,\rm GeV$ \cite{Ko:2014turic, Fries:hp2015}.

In Fig.\ref{fig:ratioppi} we compare the $p/\pi$ ratio vs $p_T$ shown by solid line
with the experimental data of the ALICE Collaboration \cite{Abelev:2014laa} shown by open
circles. The description is overall quite good with some quite limited lack of proton yield at
$p_T \sim 6 \, \rm GeV$. 
In Fig.\ref{fig:ratioppi} it is also shown by dashed line the $p/\pi$ ratio if
the coalescence between soft partons from the QGP and a mini-jet.
We can see that the contribution is significant for $p_T> 3\rm GeV$.
The impact of radial flow of the soft partons is shown by dashed double-dotted line. 

In Fig. \ref{fig:ratiolamk} we show the results for the $\Lambda/K$ ratio in comparison
with the experimental data shown by circles \cite{Abelev:2013xaa}. We can see generally
a good overall description of the ratio especially in the region of the peak.
Comparing the dashed-dot line with the thick solid line we can see 
that in the peak region a quite good agreement with the
experimental data is reached  
thanks to a recombination of thermal soft partons with a mini-jet parton. 
The relevance of such a process is present also in the EPOS approach Ref.\cite{Werner:2012sv},
ad it is an idea that can be traced back to ref.\cite{Lin:2002rw}
However as discussed above at $p_T\sim 6\,\rm GeV$ there is a significant lack of $\Lambda$
yield that here in a linear scale appears quite large. 
A tendency to underestimate the yield is visible also in \cite{Werner:2012sv}
even if quite smaller thanks to a different fragmentation scheme with respect to AKK.
The low ratio is only slightly
low because of the large K yield in this $p_T$ range,see Fig.\ref{fig:kaon}, 
and most of the disagreement with the data comes from the lack of yield in the $\Lambda's$
distribution from fragmentation that appears too soft in this $p_T$ range.
In Fig.\ref{fig:ratiolamk} we also show the behavior of the $\Lambda/K$ ratio if only
coalescence is considered, dashed line, or if only fragmentation is included.

\section{Summary and Conclusions}

In this paper, we have studied the hadronization of the quark-gluon plasma
and mini-jet partons produced in relativistic heavy ion collisions 
in terms of the parton coalescence plus fragmentation model.  
The $p_T$ distributions 
of partons in the quark-gluon plasma is taken to have an exponential 
form with a temperature similar to the phase transition
temperature, $T_c \simeq 160 \,\rm MeV$ but boosted by a radial flow $\beta$.
The volume and radial flow of the hadronizing QGP are constrained 
by the total multiplicity and total transverse energy. Partons at higher 
$p_T \simeq 15\,T_c \simeq 2.5 \,\rm GeV$ 
are taken to be  the mini-jets with power-law spectra that have undergone the jet-quenching process.
The approach is in its core the one developed at RHIC energy a decade ago \cite{Greco:2003mm}

The aim has been to investigate if the approach already successful a decade ago
with the first RHIC data was able to correctly predict the evolution of the
transverse momentum spectra from RHIC to LHC energy.
More specifically the new data also at RHIC have been available in a wider $p_T$ range
especially for the baryon-to-meson ratio and at LHC it has become possible to
verify the approach in a wider $p_T$ range up to about 10 GeV also for the $\Lambda$
and for both the $p/\pi$ and $\Lambda/K$ ratio.
We found a quite good agreement of both the $p_T$ spectra for the main hadrons
like $\pi,K, p, \bar p, \Lambda$ and the baryon-to-meson ratio $p/\pi, \Lambda/K$
in a wide region of $p_T$ up to about 10 GeV.
The yield and slope of the spectra as well as for the height and  $p_T$ 
positions of the peak in the baryon/meson ration is correctly predicted.
This is achieved without any adjustment of the coalescence parameter $\Delta_p$ for mesons
and baryons and hence can be considered as a real prediction of the model. We also note
that underlying such agreement a key ingredient is the radial flow of the QGP matter.
A coalescence model without radial flow could not account for properly neither 
for the baryon over meson
enhancement nor for the single $p_T$ hadron spectra.

With respect to the original paper \cite{Greco:2003mm} the new data extends to higher $p_T$,
especially for the $\Lambda$, and this as allowed to spot some lack of yield
for baryons in the $p_T$ region where the coalescence contribution is dying out and fragmentation
is taking over, which means at $p_T \sim 6 \,\rm GeV$. It appears that the
independent fragmentation approach gives too hard spectra
at least up to $p_T \sim 8\,\rm GeV$ especially for baryons, similarly to what has been pointed out for charged hadrons in pp collisions \cite{d'Enterria:2013vba}. 
In AA collisions this result seems to point to the need of an in-medium
fragmentation process in agreement with the first results of a 
shower recombination of quarks from gluon decays \cite{Han:2012hp,Ko:2014turic, Fries:hp2015}.

We remind that the coalescence has played an important role in the discussion about
the QGP properties not only thanks to the explanation of the baryon/meson enhancement
but also because of the approximate quark number scaling observed at RHIC and suggested
by a simplified approach to a pure coalescence mechanism \cite{Molnar:2003ff}.
A more realistic approach to coalescence in three dimension with radial flow correlations,
finite hadronic wave function widths and resonance decays shows that about a $10\%$ breaking
has to be expected at intermediate $p_T$, and quite larger one at low $p_T$ \cite{Greco:2004ex,Greco:2007nu} correctly predicting the experimental observation \cite{Adams:2004bi}.
At LHC energy experimental data show a larger breaking of the scaling with respect to
the one observed at RHIC or predicted by more realistic coalescence models.
However we note that the data discussed now are based on event-by-event analysis
that shows the presence of higher harmonics like $v_3,v_4, v_5$ which also have a quite
large variance. This can be expected to further break the naive quark number scaling
of the $v_2(p_T)$ but a quantitative approach requires an extension of the present
approach to an event-by-event Monte Carlo one. Such an approach
is currently under development and it requires a significant longer computational effort to have
a good statistical mapping of the various harmonics $v_n(p_T)$ in each event 
and in a wide range of $p_T$. However the present work, showing that also at LHC the main hadronic $p_T$ 
spectra and baryon-to-meson ratios are well predicted, provides the motivation
to extend the study to the anisotropic flows.

%\subsection{\label{sec:citeref}Citations and References}
%\subsubsection{Citations}

\section*{Acknowledgement}
The authors acknowledge the support of the ERC Grant for the QGPDyn project.
The authors are grateful to David D'Enterria, Peter Christiansen and Ron Belmont
for useful suggestions and for providing pertinent references.

%\bibliography{bib_art}{}
%\bibliographystyle{h-physrev}

\end{document}